# Compact and Robust Deep Learning Architecture for Fluorescence Lifetime Imaging and FPGA Implementation

Zhenya Zang, Dong Xiao, Quan Wang, Ziao Jiao, Yu Chen, and David Day-Uei Li

**Abstract**— This paper reported a bespoke adder-based deep learning network for time-domain fluorescence lifetime imaging (FLIM). By leveraging the $l_1$-norm extraction method, we propose a 1-D Fluorescence Lifetime AdderNet (FLAN) without multiplication-based convolutions to reduce the computational complexity. Further, we compressed fluorescence decays in temporal dimension using a log-scale merging technique to discard redundant temporal information derived as log-scaling FLAN (FLAN+LS). FLAN+LS achieves 0.11 and 0.23 compression ratios compared with FLAN and a conventional 1-D convolutional neural network (1-D CNN) while maintaining high accuracy in retrieving lifetimes. We extensively evaluated FLAN and FLAN+LS using synthetic and real data. A traditional fitting method and other non-fitting, high-accuracy algorithms were compared with our networks for synthetic data. Our networks attained a minor reconstruction error in different photon-count scenarios. For real data, we used fluorescent beads' data acquired by a confocal microscope to validate the effectiveness of real fluorophores, and our networks can differentiate beads with different lifetimes. Additionally, we implemented the network architecture on a field-programmable gate array (FPGA) with a post-quantization technique to shorten the bit-width, thereby improving computing efficiency. FLAN+LS on hardware achieves the highest computing efficiency compared to 1-D CNN and FLAN. We also discussed the applicability of our network and hardware architecture for other time-resolved biomedical applications using photon-efficient, time-resolved sensors.

---

## 1 INTRODUCTION

Fluorescence is represented by detectable photon emission from molecules or atoms, a radiative process transferring energy from an excited state to a lower state. Fluorescence lifetime is a quantitative metric to analyze fluorescence phenomena. Fluorescence lifetime imaging (FLIM) is an emerging technique boosting biomedical imaging. For example, endogenous fluorophores, such as amino acids, metabolic enzymes, and vitamins, have been engineered to assess Vivo imaging [1], [2]; exogenous fluorophores, such as Förster resonance energy transfer (FRET) pairs, have been adopted to investigate protein interactions [3] and drug developments [4]. Autofluorescence has been utilized in fluorescence-guided surgery [5], [6]. Time-correlated single-photon counting (TCSPC) [7] is the core module for time-resolved, wide-field [8], and confocal FLIM [9] as it can suppress the Poisson noise during acquisition time [10]. To guarantee precise time-of-arrival, a stop-watch employing high temporal resolution (picosecond) time-to-analogue or -digital converters can accumulate arrival photons from a train of laser pulses and generate a histogram decay. Retrieving lifetimes from temporal point spread functions (TPSF, or histogram decays of collected photons) is an ill-posed problem in which a perfect solution cannot be found. Traditional deconvolution-based nonlinear square fitting (NLSF) [11] and maximum likelihood estimation [12] methods are computationally expensive. Fast, model-free methods such as centre-of-mass [13] and its derivatives [14] were proposed to estimate lifetimes without iterations to overcome the mentioned issues. However, they are susceptible to low signal-noise ratios (SNR). They can resolve either mono- or bi-exponential decays, meaning it is hard to analyze complex fluorophores with multi-exponential decays. Moreover, they cannot simultaneously and accurately recover two critical average lifetimes, i.e. amplitude- ($\tau_A$) and intensity-averaged lifetime ($\tau_I$). The selection of the two average lifetimes should be considered for different applications [15]. This study clarified that $\tau_A$ is a crucial parameter to deduce energy transfer efficiency and dynamic quenching behaviours; $\tau_I$ is essential to solving the quenching constant from the Stern-Volmer constant. The study [16] clarified that the ratio of $\tau_A$ to $\tau_I$ is an intuitive indicator for monitoring multi-exponential fluorescence decays. Therefore, we aim to construct a model-free and robust method to reconstruct $\tau_A$ and $\tau_I$ to cover most FLIM applications. Given that generating artificial FLIM TPSFs is well-documented, data-driven approaches can boost the accuracy and computing efficiency of FLIM

---

- *Zhenya Zang, Dong Xiao, Quan Wang, Ziao Jiao, and David Day-Uei Li are with the Department of Biomedical Engineering, University of Strathclyde, G4 0RE, United Kingdom. E-mail: zhenya.zang, dong.xiao, quan.wang, ziao jiao, and david.li@strath.ac.uk.*
- *Yu Chen is with the Department of Physics, University of Strathclyde, G4 0NG, United Kingdom. E-mail: y.chen@strath.ac.uk.*

TABLE 1
COMPARISON OF DEEP LEARNING ARCHITECTURES FOR FLIM

| | 3-D CNN [31] | 1-D CNN [25] | 1-D CNN [28] | MLP [27] | Phasor-MLP [30] | MLP [29] | GAN [33] | ELM [36] | **FLAN** | **FLAN+LS** |
|---|---|---|---|---|---|---|---|---|---|---|
| Multi-exponential compatible | Yes | No | Yes | Yes | Yes | No | Yes | Yes | Yes | Yes |
| Input data type | 3-D tensor | 1-D histogram | 1-D histogram | 1-D histogram | Phasor coordinates | 1-D histogram | 1-D histogram | 1-D histogram | 1-D histogram | 1-D histogram |
| Model size | 4.14MB (1,084,045) | 0.016MB (4,248) | 0.19MB (48,675) | 0.57MB (149,252) | 1.84KB (471) | 14.31MB (3,750,250) | 0.55MbB (143,528)[a] | 0.13MB (205,600) | 0.088MB (23,003) | 0.017 MB (4,058) |
| Training time | 4 h | 17 min | 23 min | 4 hr | 15 min | 38 min | 6.9 h | 10.85 sec | 20 mins | 18 mins |
| Training hardware | GPU | GPU | GPU | GPU | CPU | GPU | GPU | CPU | GPU | GPU |
| Inferencing hardware | GPU | GPU or FPGA | GPU | GPU | CPU | GPU | GPU | CPU | GPU or FPGA | GPU or FPGA |
| Data compression | - | Quantization-aware training | - | - | - | - | - | - | Post-quantization | Post-quantization+LS |

[a] The model size was not given in the paper; the size was estimated from the network's structure.

analysis. Therefore, we constructed a compact hardware-friendly deep learning (DL) architecture for high-accuracy $\tau_A$ and $\tau_I$ reconstruction. Further, we introduced a nonlinear mapping approach to compress histograms without performance degradation.

On the other hand, an increasing number of computing platforms for biomedical devices are migrating toward the edges. For example, portable oximeters can measure oxygen concertation of flowing blood non-invasively using photoplethysmography. Such highly integrated devices perform efficient computing, yet the power budget is considerably low. Likely, we yearn to implement our DL networks on FPGA as it can achieve highly parallel, reconfigurable computing while consuming low power. We designed an automatic software script to extract, and export learned parameters from pre-trained models to FPGA, paving the way for other time-resolved biomedical applications. Further, we adopted on-chip post quantization to shorten the bit-width of the on-the-fly data. We analyzed the computing efficiency of the FPGA embedded with our neural networks and compared it with CPU and GPU.

The rest structure of the paper is that Section 2 reviews and summarizes the existing FLIM algorithms on hardware; Section 3 mathematically clarifies the optical characterization of FLIM and explains lifetime parameters to be revolved; Section 4 demonstrates the network architecture and training details; Section 5 describes the data compression strategies; Section 6 and 7 evaluate our networks using synthetic and real data, respectively; Section 8 reveals details of networks' FPGA implementation; Section 9 conclude this work and discuss the potential improvements.

## 2 PRIOR WORK
### 2.1 Hardware platforms for FLIM
Graphics processing units (GPUs) have been extensively adopted as hardware accelerators for FLIM. Byungyeon *et al.* [17] proposed a mean-delay technique and implemented it on a GPU to directly extract fluorescence lifetime from analogue fluorescent decays. Additionally, a GPU was utilized to realize a pixel-wise, fitting-free phasor method [18] towards a two-photon FLIM system to achieve video-rate fluorescence lifetime inference. A compressed-sensing algorithm (CS) was applied to a high-resolution, wide-field FLIM system that used compressed ultrafast photography [19]. A GPU-based computer cluster was used to solve the iterative reconstruction algorithm. An FPGA is the ideal hardware for data readout and computing due to its sufficient I/O interfaces and reconfigurable computing parallelism. Particularly, single-photon avalanche diode (SPAD) arrays [20] coupled with FPGA as parallel and high-throughput data readout are needed. Therefore, some FPGA-friendly FLIM algorithms were proposed. Embedded center-of-mass (CMM) algorithms were implemented on FPGAs [21] and application-specific integrated circuits (ASICs) [22], [23] to determine lifetimes from time-to-digital converters' time stamps directly. Another CMM processor was implemented on an FPGA coupled with a SPAD-based silicon photomultiplier sensor [24], realizing real-time fluorescence lifetime estimation and online fluorescence lifetime cytometric sorting. After that, a DL prototype that is less susceptible to noise [25] was embedded on an FPGA to estimate fluorescence lifetime in a fluorescent flow cytometry system. On-chip lifetime estimation was also realized in the frequency domain; discrete Fourier transform was embedded on an FPGA to compute phase lifetimes and modulations [26].

### 2.2 Deep Learning for FLIM
In the recent decade, DL has been an emerging approach for FLIM analysis. An MLP [27] was initially reported to analyze bi-exponential decays for a two-photon system. To build a noise-robust DL architecture, a 1-D convolutional neural network (CNN) [28] was proposed to resolve bi- and tri-exponential decays with low SNRs. An MLP [29] was applied on a high spatial resolution, wide-field FLIM using time-gated SPAD without raster-scanning to miniaturize the FLIM system. Another MLP [30] processes phasor coordinates and generates lifetime parameters. Because the inputs are coordinates rather than histograms, the MLP exhibits fast speed. To extract both spatial and temporal features from raw fluorescence data tensors, a 3-D CNN [31] was reported for near-infrared FLIM applications. Besides, Ruoyang *et al.* proposed a 3-D CNN combined with CS [32] for wide-field FLIM imaging *in vitro* and *in vivo* environments. Further, to address the challenge of FLIM with photon-starved conditions, Yuan-I *et al.* reported a generative adversarial network [33] for FLIM imaging in low-photon scenarios (below 400). Another challenge of FLIM is the spectral overlap of fluorescent emissions leading to bleedthrough. Jason *et al.* proposed a hybrid CNN architecture [34] to extract individual lifetime components from multi-exponential, hyperspectral fluorescence emission decays. To enhance the spatial resolution of FLIM images, Dong *et al.* [35] introduced a cascade CNN architecture to infer fluorescence lifetime and improve the spatial resolution afterwards. An Extreme Learning Machine (ELM) [36] was presented to achieve fast and accurate lifetimes reconstruction with back-propagation-free, online training.

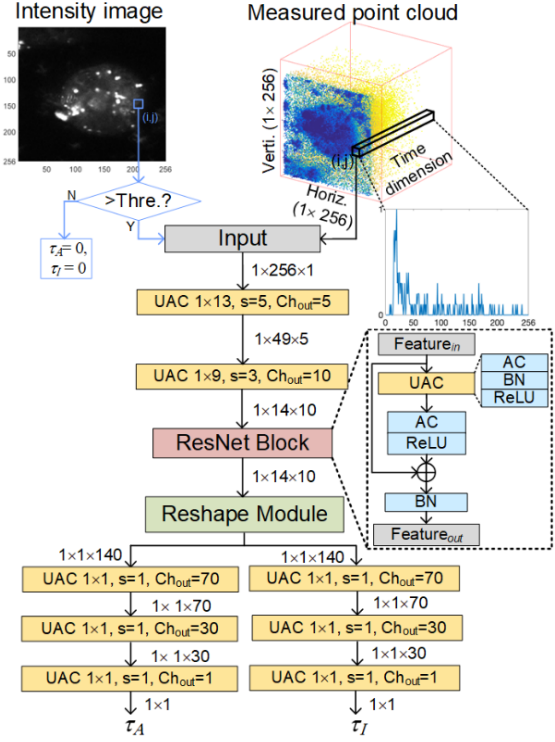

Fig. 1. Overview of FLAN architecture. Histograms from each pixel will be processed consecutively, and the pixels from the background will be discarded initially. Prostatic cells coated with bio-markers are adopted here for illustration.

## 3 PROBLEM DEFINITION

Our initial objective is to retrieve lifetime parameters from TPSFs, where TPSF $h(t)$ for each focal point can be formulated

**Algorithm 1** Unified AC for each histogram

**Input:** input FM $F_i$ of shape $CH_i \times W_i$;
filters $W$ of shape $K_x \times CH_i \times CH_o$;
BN coefficients $scale$ and $shift$ of shape $CH_o$;
stride $S$; Threshold $T$; photon count $N_{pc}$;
**Output:** output FM $F_o$ of shape $CH_o \times W_o$;
1: **if** $N_{pc} > T$ **then**
2:   **for** $ch_i = 0; ch_i < CH_i; ch_i{+}{+}$ **do**
3:     **for** $w_o = 0; w_o < W_o; w_o{+}{+}$ **do**
4:       **for** $k_x = 0; k_x < K_x; k_x{+}{+}$ **do**
5:         $w = ch_i \times S + k_x$;
6:         **if** $(w >= 0)\&\&(w <= W_i)$ **then**
7:           **while** $ch_o < CH_o$ **do**
8:             $F_o[w_o][ch_o] \mathrel{+}= -|F_i[w][ch_i]$
              $-W[k_x][ch_i][ch_o]|$;
9:             $ch_o{+}{+}$;
10:   $F_o = F_o \times scale[ch_o] + shift[ch_o]$;
11:   $F_o = ReLU(F_o)$;
12: **else**
13:   $F_o = 0$;

by the convolution of the periodic IRF $I(t)$ and the multi-exponential probability density functions (PDF) $p(t)$

$$h(t) = I(t) * p(t) + \rho(t) \\ = I(t) * \left[A \sum_{i=1}^{N} a_i \exp(-t/\tau_i)\right] + \varepsilon(t), \quad (1)$$

where $\sum_{i=1}^{N} a_1 = 1$, $A$ depicts the amplitude, $a_i$ and $\tau_i$ means the amplitude fractions and lifetime of an individual lifetime component, $N$ implies the number of lifetime components, and $\varepsilon(t)$ is the Poisson noise. The intensity-averaged lifetime $\tau_I$ means the average lifetime a fluorophore stays excited, depicted by

$$\tau_I = \frac{\int_0^\infty t \sum a_i \exp(-t/\tau_i) dt}{\int_0^\infty \sum a_i \exp(-t/\tau_i) dt} = \frac{\sum_{i=1}^{N} a_i \tau_i^2}{\sum_{i=1}^{N} a_i \tau_i}. \quad (2)$$

An amplitude-averaged lifetime is the integral of an intensity decay that is equivalent to steady-state intensity, formulated by

$$\tau_A = \int_0^\infty \sum a_i \exp(-t/\tau_i) dt = \sum_{i=1}^{N} a_i \tau_i. \quad (3)$$

## 4 DEEP LEARNING NETWORK DETAILS

We intend to achieve high-fidelity FLIM images with a minor computing overhead. The extreme case of lightweight DL networks is the binarized-neural network [37] that only involves bit-wise operations. However, it suffers from low accuracy in practical applications [38]. It was proved that 2-D AdderNet [39] could achieve comparable accuracy with CNN without matrix multiplications. Here, we introduce a derivative of the original AdderNet, compressed 1-D fluorescence lifetime AdderNet (FLAN), catering to the histogram decays' processing.

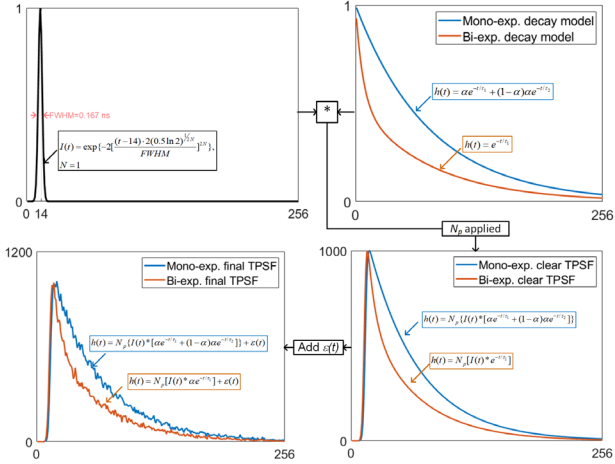

Fig. 2. Procedures for generating mono- and bi- TPSFs

## 4.1 Network Architecture

The naïve AdderNet was initially used for 2-D image classification [39]. It has been proved that adder-convolutions (AC) using $l_1$ distance to calculate similarities between feature maps (FM) and kernels can achieve nearly identical accuracy versus traditional convolutions. We use strides in each AC instead of zero-padding to simplify hardware logic to enlarge reception fields. The schematic of FLAN is shown in Fig. 1. We constructed unified AC (UAC) modules including AC, ReLU and batch-normalization (BN) to facilitate hardware implementation. And a ResNet [40] block was placed in the network's backbone to alleviate gradient vanishing. Differentiating from our previous work [28] that extracting individual lifetime components, we adopted a branched topology to infer two average lifetimes $\tau_A$ and $\tau_I$ that can analyze most FLIM applications. Algorithm 1 demonstrates the entry and the first layer of FLAN. A predefined threshold ($T$) was applied to filter out background pixels (Line 1). If the pixel is not from the background ($N_{pc} > T$), a UAC will process the histogram (from Line 2 to 11). Moreover, we proposed FLAN+LS that compresses histograms using log-scale nonlinear mapping. FLAN+LS needs fewer down-sampling layers because the length of input histograms was compressed in advance. Therefore, FLAN+LS is more compact and saves less computing overhead than FLAN. The details of compression will be illustrated in Section 5.

We compared the size of existing DLs for FLIM, and the compression ratio is defined by the ratio of FLAN+LS's parameter size to uncompressed models. Table 1 summarizes current DL architectures for FLIM. Among the existing architectures that can resolve multi-exponential decays, our models obtain the smallest parameter size apart from Phasor-MLP [39]. Phasor-MLP has the fewest parameters as histograms were initially converted into phasor coordinates using the Fourier transform. Despite the simplicity, reconstructed lifetime parameters sometimes had distinct dispersion than 3-D CNN [31]. ELM [36] has the simplest structure containing only one hidden layer.

In contrast, the hidden layer contains 500 nodes. ELM's computational complexity exponentially increases as the number of nodes grows. Such a complex, fully connected structure is not efficient for hardware implementation. Besides, although the 1-D CNN [25] for fluorescence flow cytometry has a compact size and was optimized and implemented on FPGA, it cannot analyze multi-exponential decays. As the data throughput of PMT or SPAD array is enormous (Gigabyte per second), lightweight DL architecture and efficient data compression are imperative to alleviate the bandwidth of the data transfer. FLAN and FLAN+LS can efficiently address the issues with the ACs and histogram compression method.

## 4.2 Preparation for Hybrid Training Data

As bi-exponential models can well approximate multi-exponential decays [16]. We generated hybrid datasets containing mono- and bi-exponential decays according to (1). The generation flow is shown in Fig. 2. The synthetic IRF with 0.167 ns Full width at half maximum (FWHM) was simulated following a Gaussian curve centre at the 14[th] time bin that is consistent with our two-photon system, given by

$$I(t) = \exp\left\{-2\left[\frac{(t-14)2(0.5\ln 2)^{1/2N}}{FWHM}\right]^{2N}\right\}, \quad (4)$$

where we set $N = 1$ to make the synthetic IRF close to the real IRF. The mono- and bi-exponential PDFs were generated using the same instrumental parameters. Then bespoke TPSF were generated by convolving the IRF with PDFs. Poisson noise was added, ultimately using *Poissrnd* () in Matlab to mimic background and shot noise. Peak photon counts ($N_p$) and lifetime parameters ($\tau$ for mono-exponential, $\tau_1$, $\tau_2$, and $\alpha$ for bi-exponential decays) were randomly selected in given ranges, as shown in Table 2. Each synthetic TPSF was labelled with GT lifetime parameters to train our networks afterwards. Then, these two types of TPSFs were shuffled to finalize the datasets.

## 4.3 Training Details

FLAN and FLAN+LS architectures were implemented by PyTorch and trained on a 16 GB NVIDIA Quadro RTX 5000 GPU. To accelerate the training, we employed adaptive learning rates (ALR) with an initial value of 0.001 and a momentum factor of 0.995. RMSProp is the optimizer in the training phase. 50,000 synthetic decays were generated for training purposes; extra 5,000 decays were utilized for validation. Early-stopping with 20-epoch patience was adopted to prevent overfitting. ReLU is the activation function after each AC layer. BN was used to mitigate the internal covariate shite of FMs in the network, whereby the training can be accelerated. The adopted loss function is MSE, depicted as

$$\mathcal{L}(\Theta) = \frac{1}{B_1} \sum_{b_1}^{B_1} \sum_{b_2}^{B_2} \left\| F_{b2}(h^{b_1}, \Theta) - G^{b_1} \right\|_2^2, \quad (5)$$

where $B_1$ and $B_2$ are the batch size (128) and the number of output nodes (2), $F(\cdot)$ is the end-to-end inference function of training parameters $\Theta$, and $G$ is the set of ground truth lifetimes. To make comparisons, we re-trained the 1-D CNN [28] with traditional multiplications-based convolutional kernels for comparison, where we kept the networks' topology identical to FLAN and FLAN+LS. As the learning rate (0.0001) of the original 1-D CNN is constant, it would exhibit a slower convergent speed than ALR. The loss curves of the three networks are shown in Fig. 3. FLAN+LS achieves a similar loss curve versus FLAN even with fewer time bins. And these two networks obtained similar ending loss versus 1-D CNN when training ceased.

TABLE 2
RANGES OF LIFETIME PARAMETERS FOR SYNTHETIC DATA

|  | $\tau_1$ (ns) | $\tau_2$ (ns) | $a$ | $\tau_I$ (ns) | $\tau_A$ (ns) |
|---|---|---|---|---|---|
| Mono-exp | [0.1, 5] | - | 1 | [0.1, 5] | [0.1, 5] |
| Bi-exp. | [0.1, 0.5] | [1, 3] | [0, 1] | [0.01, 3.98] | [0.01, 3.98] |

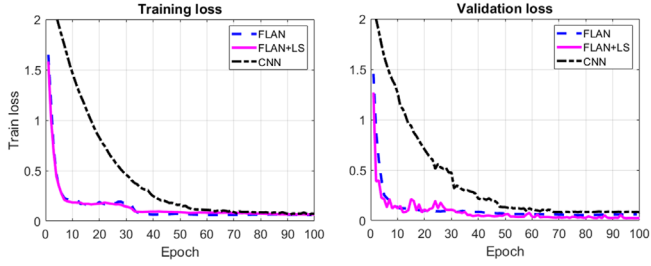

Fig. 3. Training and validation loss curves from 1-D CNN, FLAN, and FLAN+LS.

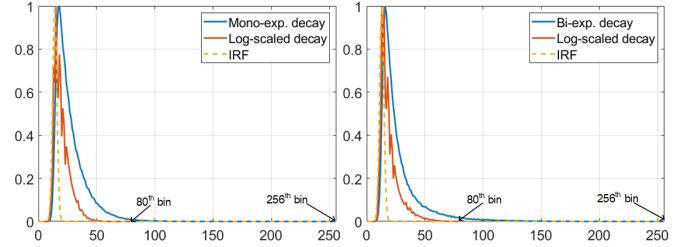

Fig. 4. IRFs, original and compressed synthetic decays. (a) Mono-exponential decay with lifetime. (b) Bi-exponential decay with lifetime.

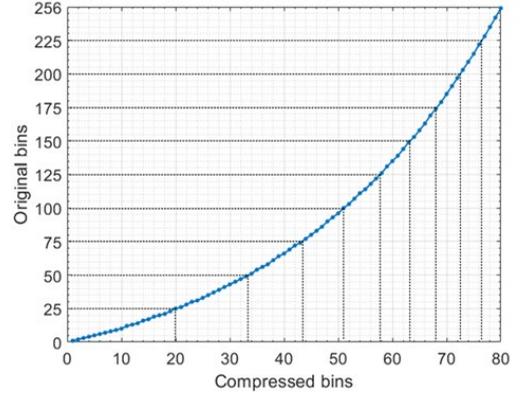

Fig. 5. Nonlinear time-bin mapping, merging original 256 time-bin to compressed 80 time-bin.

## 5 DATA COMPRESSION STRATEGIES

We introduce two-fold data compression methods to alleviate computing latency and reduce data movements. At first, we compressed histograms using a log-scale bin-merging technique to reduce redundant time bins. And then, a post-linear-quantization was employed on FPGA to compress bit-width throughout the FLAN and FLAN+LS architecture.

### 5.1 Log-scale Mapping for Time-bin

The fluorescence decays are less informative as the index of time-bin increases, especially for the specimen with ultra-small lifetimes. As blue lines show in Fig. 4, they are mono-and bi-exponential decays with lifetimes of 0.5 ns and 0.45 ns, respectively. No photon occurred after 86[th] and 84[th] time-bin, meaning that time bins behind them can be merged to compress the histogram and moderate the sparsity. Therefore, the computational complexity can be decreased. The target number of time bins $M$ was controlled by a factor that should be pre-defined; the time-bin interval of the compressed histogram is

$$S = s(x+1) - s(x) = \\ floor(\frac{r^x - 1}{r - 1}) - floor(\frac{r^{x+1} - 1}{r - 1}), \quad (5)$$

where $floor\,(\cdot)$ rounds each element to the nearest integer less than or equal to that element, $T$ is the original number of time-bin (256), $x$ is the index of compressed time-bin. $r$ is calculated from $M$, given by

$$\frac{r^M - 1}{r - 1} = T \quad (6)$$

where $M$ is 80 in our experiments because the compressed histograms can reserve sufficient temporal information after merging. The number of photons in the new time interval is the summation of the counts from the original intervals. The nonlinear mapping process is depicted in Fig. 5; more temporal information can be reserved for the front bins. Accordingly, in Fig. 4, both typically synthesized mono- and bi-exponential decays with 256 time-bin can be effectively converted into 80 bins. Compressed decays with fewer time-bin will simplify the neural network. For the decays with greater lifetimes, $M$ can be easily adjusted; and the training data can be generated accordingly.

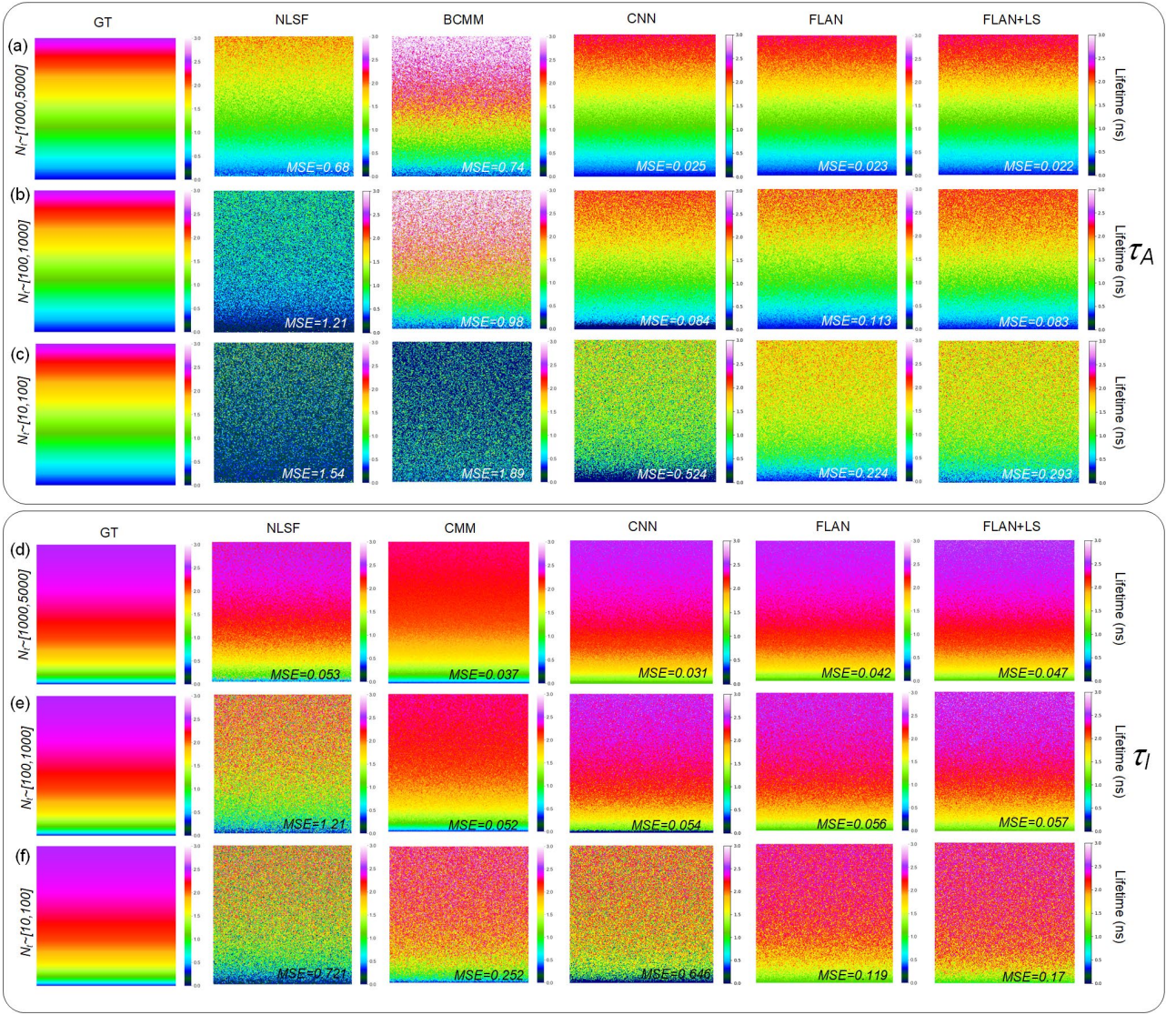

Fig. 6. Synthetic GT and reconstructed $\tau_A$ and $\tau_I$ FLIM images. MSE is used to evaluate the accuracy. (a), (b), and (c) reconstructed $\tau_A$ in [1000, 5000], [100, 1000], and [10, 100] photon-count. (d), (e), and (f) reconstructed $\tau_I$ in [1000, 5000], [100, 1000], and [10, 100] photon-count.

### 5.2 On-chip Linear Quantization

Quantization for FMs and learned parameters are well-known approximate arithmetic for DL. And it is economical for logic utilization and power consumption on FPGA. We adopted the linear quantization (or static fixed-point) strategy [41] to convert floating-point (FLP) learned parameters and FMs in pre-trained models into fixed-point (FP). To maintain the computing accuracy, we configured FMs with 16-bit integer and 16-bit fractional length segmentation; learned parameters with 10-bit integer and 10-bit fractional parts. The conversion can be depicted as

$$\begin{cases} D = round(N_{FLP} \times 2^F) \\ B = D2B(B) \\ N_{FL} = shift(B, F) \end{cases}, \qquad (7)$$

where $F$ is the fractional length of the FP number, $D2B(\cdot)$ means converting from decimal to binary type, $shift(\cdot)$ moves the decimal three places to the left, and $N_{FLP}$ and $N_{FP}$ depict the FLP number and binarized FP number, respectively. The precision of the FP number is $2^{-10}$. The quantization can be realized using FP libraries in Vivado High-Level-Synthesis (HLS).

## 6 SYNTHETIC DATA EVALUATION

With hybrid test datasets containing shuffled mono- and bi-exponential decays, we quantitatively evaluated FLAN and FLAN+LS in terms of $\tau_A$ and $\tau_I$ reconstruction. And we compared them with traditional curve fitting and other high-accuracy algorithms in different photon-count conditions. To generate intuitive 2-D GT FLIM images, two individual lifetime components ($\tau_1$ and $\tau_2$ in (1)) were fixed values of 0.3 ns and 2.5 ns. And from the top to the bottom of GT lifetimes, amplitude fractions uniformly increased from 0 to 1. Mean square error (MSE) was used to evaluate the accuracy of each method in Fig. 6. Three deep learning methods achieved better recovery performance for both $\tau_A$ and $\tau_I$. And both FLAN and FLAN+LS outperformed 1-D CNN under different photon-count conditions with the same training datasets. Despite the compression of FLAN+LS, the performance was not deteriorated, even in low-count conditions. In terms of $\tau_A$, the performance of NLSF declined as photon counts decreased because the fitting process (Levenberg-Marquart deconvolution) required enough counts to guarantee accuracy. The provocative study [42] using similar synthetic parameters to ours unravelled that, for mono-exponential decays, the necessary number of photons of the deconvolution method is around 100 for the minimal resolvable lifetime of 0.3 ns. Moreover, pre-set initial values are critical for the deconvolution method, and we did not fine-tune them here. BCMM was also susceptible to the low-count condition (below 100) as Romberg's integration in BCMM is inaccurate for sparse vectors. As for $\tau_I$, in Fig 6 (d) and (e)., although CMM achieved smaller MSE than deep learning methods under high and medium photon counts conditions, discernable bias can be observed in the top half of CMM's recovered images. Before using CMM, prior knowledge about the width of the analysis window is required to achieve optimal results. To observe the lifetime distributions more efficiently, we visualized the lifetime distributions of all recovered and GT images in Fig 7.

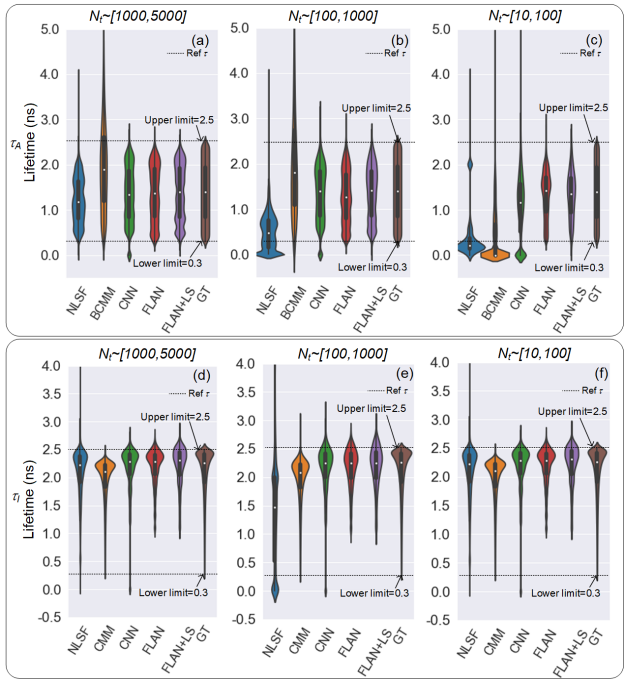

Fig. 7. Lfetimes' distributions of $\tau_A$ in (a), (b), and (c); $\tau_I$ in (d), (e), and (f) retrieved from different algorithms in different levels of photon count.

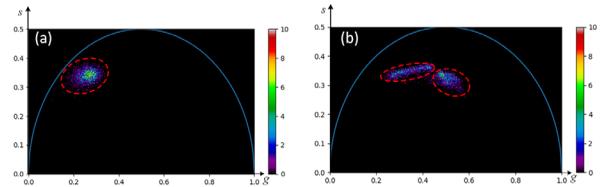

Fig. 8. Phasor plots of unmixed (Yellow-green and Crismon) and mixed (Yellow-green) fluorescent beads, showing (a) one cluster and (b) two clusters.

## 7 REAL CASE STUDY: FLUORESCENT BEADS DISCRIMINATION

To evaluate our networks using fluorescent samples, we used different algorithms to retrieve lifetimes based on fluorescent beads with characterized lifetimes. This section introduces the sample preparation processes, optical setup, and evaluation results.

### 7.1 Sample preparation

We chose fluorescent latex beads to evaluate the accuracy of our networks as they are intuitive tools applied to tracing fluid dynamics [43] and validating hyperspectral imaging microscopy [44]. Here, yellow-green and crimson fluorescent beads of 10 μm diameter (F8831, FluoSphere Polystyrene Microspheres, Thermo Fisher, UK) were dissolved in an aqueous concentration of $3.6 \times 10^5$ beads/ml. We dropped the diluted solution of fluorescent beads onto a microscope slide covered with coverslips and then mounted it on microscope sample holder of a two-photon system. The unmixed set encompasses yellow-green beads with a reference lifetime of 2.1 ns. Alone with the unmixed group, the mixed set includes extra crimson beads with a 3 ns reference lifetime. The reference lifetime was measured using the TCSPC technique on a Horiba Deltaflex fluorometer and analysized using Horiba DAS6 software.

### 7.2 Optical Setup

The utilized excitation source was a femtosecond Ti: sapphire laser (Chameleon, Coherent, Santa Clara, USA) with tunable wavelengths, and the laser pulse width is less than 200 fs with an 80 MHz repetition rate. The emission light was collected by a 10× objective lens (N.A = 0.25) with a 685 nm short pass filter. Our setup acquired both sets of beads using LSM 510 confocal laser scanning microscope. A TCSPC (SPC-830, Becker & Hickl GmbH) card was mounted on the computer with a PCIe interface to record the time-of-flight of fluorescence emission. The spatial resolution of the FLIM data was configured as 256x256 on the SPImage software, and the field of view was selected by moving the translation stage. The number of time-bin was configured as 256 in each pixel, and 39.06 ps was the temporal resolution in each time-bin.

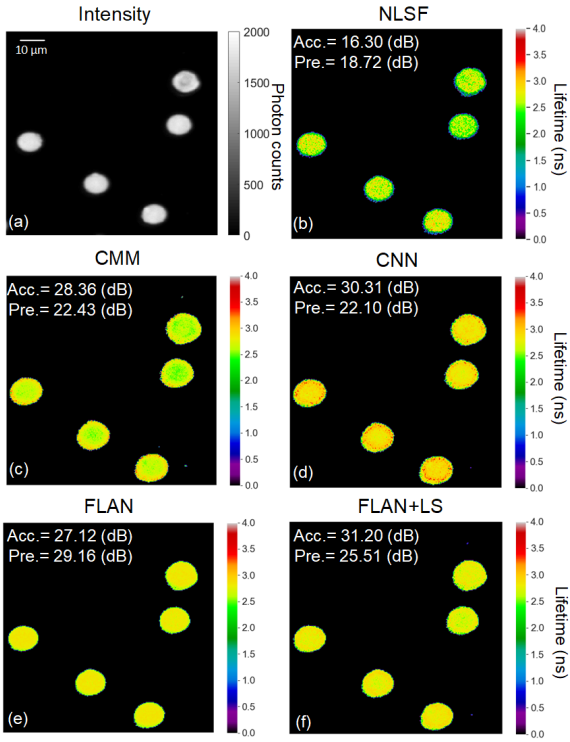

Fig. 9. Unmixed beads (Yellow-green) evaluation using accuracy and precision.

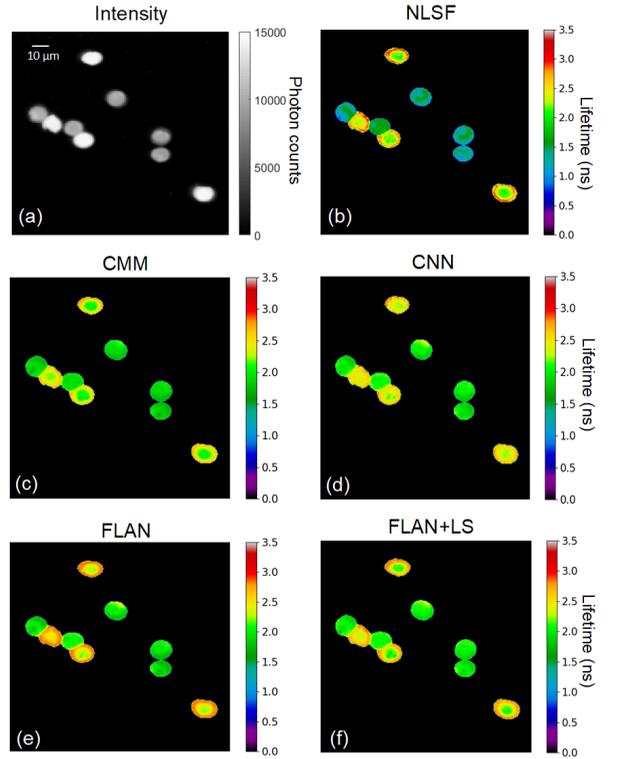

Fig. 10. Lifetime reconstruction of unmixed beads (Yellow-green and Crismon).

### 7.3 Quantitative Analysis

Given that phasor [45] is a fitting-free, frequency-domain approach to observing lifetime distributions represented by polar coordinates $g$ and $s$

$$\begin{cases} g_{i,j} = \dfrac{\int_0^\infty h(t)_{i,j} \cos(kwt)dt}{\int_0^\infty h(t)_{i,j} dt}, \\ s_{i,j} = \dfrac{\int_0^\infty h(t)_{i,j} \sin(kwt)dt}{\int_0^\infty h(t)_{i,j} dt} \end{cases} \quad (8)$$

where $i$ and $j$ is the pixels' index in the FLIM image, $kw$ is the angular frequency of the excitation source. Here, we initially used a phasor plot to analyze distributions of lifetimes in each pixel to validate the prepared beads. As shown in Fig. 8, the cluster is not located in the semi-circle, meaning that the two types of beads are not perfect mono- and bi-exponential decays resulting from a mixture of dyes. Given that there is no spectral overlap of the beads, meaning no energy transfer occurs, we used $\tau_I$ to analyze reconstructed lifetime images and to compare different algorithms. Two metrics, i.e. accuracy and precision in $dB$ [25], were used for evaluation, defined as

$$\begin{cases} Acc. = 20 \log_{10} mean(\hat{\tau}_I / \Delta\tau_I), \\ Pre. = 20 \log_{10} mean(\hat{\tau}_I / \sigma\tau_I) \end{cases} \quad (9)$$

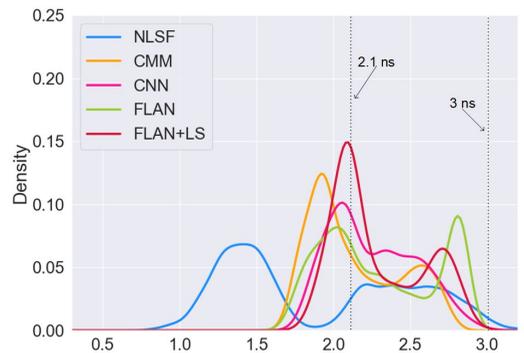

Fig. 11. Reconstructed lifetime distributions of mixed beads using different algorithms. The reference lifetime of Yellow-green and Crimson are 2.1 ns and 3 ns, respectively.

where $\hat{\tau}_I$ is reconstructed lifetimes, $\Delta\tau_I$ and $\sigma\tau_I$ means absolute error and standard deviation of the reconstructed lifetimes. From Fig. 10, we can tell DL methods obtained similar accuracy versus CMM, yet FLAN and FLAN+LS exhibit the best precision (small standard deviations). Note that NLSF has an obvious bias compared to others because it involves iterative deconvolutions that significantly rely on initial values. Likewise, we used $\tau_I$ to evaluate the algorithms for mixed beads. According to Fig. 8 (b), as there were two individual clusters, we applied different algorithms to discriminate between these two types of beads. Reconstructed FLIM images and lifetime distributions from each algorithm were plotted in Fig. 10 and 11. The CNN is less discerning than FLAN and FLAN+LS to discriminate between the two types of beads. CMM can distinguish the beads well, but bias still exists. Peaks' indexes of FLAN and FLAN+LS are closest to the reference lifetime. Hence, our networks resolve two types of beads with accuracy lifetime estimation.

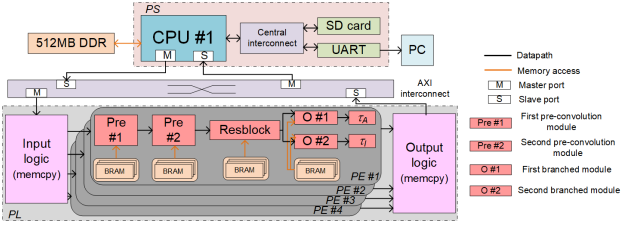

Fig. 12. Overview of the FPGA hardware platform embedding four FLANs.

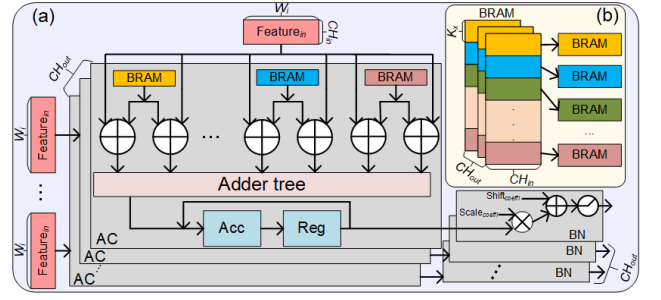

Fig. 13. (a) Architecture of UAC. BRAMs storing and caching learned parameters and FMs are partitioned to feed data to corresponding ACs. Multiple ACs and BN modules are instantiated to improve parallelism. (b) BRAMs were partitioned into smaller portions to cache parallel data on-the-fly.

## 8 Hardware Implementation

The FLAN and FLAN+LS were implemented on a PYNQ-Z2 board integrating the Processing System (PS aka ARM Cortex-A9-based CPU) and the Programmable logic (PL aka Programmable Logic). The DL cores were implemented in the PL part using HLS. Hardware drivers of the cores and less computing-intensive arithmetics, such as eliminating background pixels and decoding input FLP histograms into FP, were programmed in the PS. As shown in Fig. 12, four DL cores were instantiated to process four pixels concurrently to achieve instances-level parallel computing. The four consecutive histograms from an AXI- bus (through a high-performance slave port) should be copied into four memory buffers. And then, the four chunks of data will be fed into corresponding DL cores. Learned parameters, including weights and BN, were fetched from the pre-trained model and exported into BRAMs using an automatic Python script. Thanks to the quantization strategy, all the parameters encoded in shortened bit-width save memories and can be pre-stored in on-chip BRAMs instead of off-chip DDR memory, which decreases the latency of data movement. Before computing lifetimes, according to the pre-defined threshold, a binarized mask map with zeros and ones should be generated offline. Only the pixels' indexes with ones will be processed. Similar to learned parameters, original input histograms should also be encoded from FLP to FP. This encoding process was conducted in the ARM processor. Eventually, the output lifetimes will be converted to readable FLP after lifetimes inference. As mentioned in Section IV A, gathered AC, ReLU, and BN can simplify the hardware implementation; we merge the three modules into one hardware module, indicated by grey areas next to the main AC modules in Fig 13 (a). The original channel-wise

BN operation in $i^{th}$ output channel of forwards-propagation is

$$x_{BN} = \frac{(x(i) - \mu(i))}{\sqrt[2]{\sigma(i)^2 + \varepsilon}} \gamma(i) + \beta(i), \quad (10)$$

where $\gamma$ and $\beta$ are learned parameters that can be parsed from pre-trained models, $\mu$ and $\sigma^2$ are the statistical mean and standard deviation of $x$, $\varepsilon$ is a constant to avoid dividing by zero (0.0001 by default). The scaling and shift coefficients can be extracted as

$$\begin{cases} scale(i) = \dfrac{\gamma(i)}{\sqrt[2]{\sigma(i)^2 + \varepsilon}} \\ shift(i) = \beta(i) - \dfrac{\gamma(i)\mu(i)}{\sqrt[2]{\sigma(i)^2 + \varepsilon}} \end{cases} \quad (11)$$

Therefore, BN can be simplified as matrix multiplications and vector additions on the hardware.

$$x_{BN} = scale(i) \times x(i) + shift(i), \quad (12)$$

$scale$ and $shift$ were calculated and extracted offline using a Python script. As limited writing and reading ports of BRAM would hinder simultaneous data transfer to multiple PEs. To overcome the challenge, BRAMs under usage were partitioned into small portions to increase write-read bandwidth, leading to more hardware consumption. Due to our models' compaction, all the intermediate results among layers were cached in on-chip BRAM instead of accessing the DDR memory. After processing four adjacent pixels, only a vector containing eight inferred lifetime values will be output to ARM via address-mapping AXI for anti-quantization decoding and validation. Table 3 compares hardware utilization and computing performance of three pixel-wise DL networks, namely, 1-D CNN with 32-b FLP, FLAN with 16-b FP, and FLAN+LS with 16-b FP, implemented on FPGA. The accuracy was assessed by MSE using synthetic test datasets. We randomly picked 100 histograms and divided them into 25 batches as our DL processors can process 4 pixels concurrently. And the batches of histograms were stored in the DDR. For 1-D CNN, only 2 DL cores can be accommodated due to the FLP data format, whereby the 100 histograms should be divided into 50 batches. The latency of the three networks is presented by the consumed time for processing 4 pixels. Throughput was defined by processed pixels per millisecond (PMS). MSE was adopted to assess the accuracy, aligning with Fig. 6. The hardware utilization was obtained after place-and-route implementation. The power consumption was estimated in the post-implementation phase using power analyzer tools in Vivado 2018.2. FLAN and FLAN+LS with shorter bit-width save multitudes of hardware and achieve high operating frequency. FLAN and FLAN+LS still utilize DSPs as multiplications of BN were synthesized using DSP. Although four DL units were implemented in FLAN+LS, it consumes the fewest look-up-table (LUT) and flip-flops (FF). Due to the numerous partitioned memory blocks in FLAN and FLAN+LS, more distributed (LUT) RAMs were consumed to meet the memory bandwidth. Despite the on-chip quantization, there was no apparent accuracy degradation for $\tau_A$ and $\tau_I$ reconstruction.

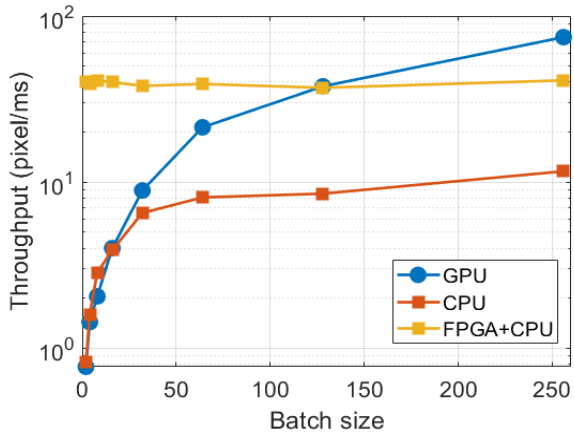

Fig. 14. Relationships between batch sizes and throughput (pixel/ms) in the inference phase on CPU, GPU, and FPGA+CPU.

TABLE 3
COMPARISONS OF HARDWARE-EFFICIENT DL NETWORKS ON FPGA

|  | 1-D CNN (2-core) | FLAN (4-core) | FLAN+LS (4-core) |
|---|---|---|---|
| Frequency | 98 MHz | 110 MHz | 120 MHz |
| Precision | 32-b FLP | 16-b FP | 16-b FP |
| DSP (220 Ava.) | 51 (23.2%) | 100 (45.4%) | 76 (34.6%) |
| LUT (53.2K Ava.) | 51072 (95.8%) | 25886 (48.6%) | 18553 (34.9%) |
| FF (106.3K Ava.) | 30827 (28.7%) | 18816 (17.7%) | 15629 (14.7%) |
| BRAM (140 Ava.) | 41 (29.1%) | 124.5 (88.9%) | 47 (33.6%) |
| LUTRAM (17.4 K Ava.) | 1566 (9%) | 2202 (12.7%) | 2198 (12.6%) |
| $\tau_A$ Accuracy (MSE) | 0.121 ns | 0.125 ns | 0.138 ns |
| $\tau_I$ Accuracy (MSE) | 0.089 ns | 0.083 ns | 0.102 ns |
| Power (W) | 2.56 | 1.93 | 1.78 |
| Latency (ms) | 1.73 | 0.16 | 0.11 |
| Throughout (PMS) | 15 | 33 | 40 |
| Efficiency (PMS/W) | 5.86 | 17.10 | 22.47 |

We also investigated how batch sizes would affect the performance of FLAN+LS on CPU, GPU, and FPGA, depicted in Fig. 14. GPU's computing performance significantly relied on the batch size and the scale of DL models. Small batch size and model size would dramatically deteriorate the speed. The figure shows that CPU and GPU present similar throughput when the batch size is smaller than 32. GPU's throughput exceeds CPU when the batch size exceeds 32 and FPGA when the batch size is more significant than 128. Note that, for some FLIM microscope applications like ours, the contrast between the sample and background should be distinguished, whereby each pixel should be compared to the threshold. This compels GPU process histograms pixel-by-pixel in serial (batch size equals 1) instead of parallel. Moreover, pixel-wise data movement on a GPU consumes extra time.

Another reason hindering GPU's performance is that CUDA and cuDNN are not providing optimizations for adder-based convolutions. As for FPGAs, they are not restricted by batch sizes, and they excel at processing data streams and addition operations.

## 9 CONCLUSION AND OUTLOOK

This study reports a multiplication-free and robust DL network for time-domain FLIM analysis. Besides, two data compression strategies were introduced to accelerate the processing. Both synthetic and real data were adopted to extensively and quantitatively demonstrate superior results comparing existing algorithms. Besides, to illustrate the applicability of hardware, we implemented our DL architectures on FPGA and compared them with traditional 1-D CNN on hardware. Results show that our compression methods can significantly save hardware resources and achieve high computational efficiency without performance deterioration. Also, to investigate our lightweight DL network's performance on GPU, CPU, and FPGA+CPU. We conclude that our heterogenous FPGA+CPU is not affected by batch sizes and outperforms GPU for small batch sizes.

A potential improvement of this work is that the training datasets and the network architecture are tailored for confocal microscopes that require a scanning mechanism and long acquisition time to guarantee high SNR of the captured data. Whereas for wide-field FLIM, imaging systems using SPAD array do not use scanning strategies but wide-field illumination, leading to shorter acquisition time and low SNR. And IRFs vary for different systems. Therefore, training datasets and the IRF should be modified to cater to various single-photon sensors. Further, our DL hardware cores cannot be integrated with the TCSPC module coupled with the confocal microscope due to the non-configurable property of the TCSPC board. However, it is possible to embed the DL cores in the data readout circuit of SPAD arrays because FPGA acts as the readout and control hardware for SPAD arrays. We will make our DL more generalized for SPAD array-based systems in software and hardware aspects.

Apart from FLIM, the DL network can be applied to analyze data from diffuse correlation spectroscopy (DCS) [46] that also relies on single-photon sensors and TCSPC systems. For DCS, the essential task is to retrieve scattering coefficients and blood flow index from temporal correlation functions. To migrate the workflow in this work to DCS, we need to alter the theoretical modelling of training datasets and the optical path, and this work paves the way for DCS research.

## ACKNOWLEDGEMENT

This work is supported by Datalab, Photon Force Ltd., and Medical Research Scottland and partly by the Biotechnology and Biological Sciences Research Council (BB/V019643/1).

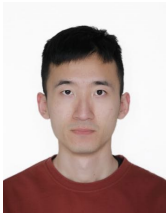

**Zhenya Zang** received an MSc degree in electronic information engineering from the City University of Hong Kong, Hong Kong. He is working toward a Ph.D. degree funded by Photon Force Ltd and Datalab, Edinburgh with the University of Strathclyde, U.K. His research mainly focuses on computational imaging for biological applications, deep learning, and reconfigurable VLSI design.

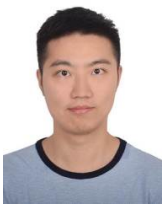

**Dong Xiao** received the M.S.degree in optics from Shenzhen University, Shenzhen, China, in 2016. He is currently working toward a Ph.D. degree funded by Medical Research Scotland and Photon Force, Ltd., Edinburgh with the Strathclyde Institute of Pharmacy and Biomedical Sciences, University of Strathclyde, Glasgow, U.K. His current research interests include fluorescence lifetime imaging microscopy systems, flow cytometry, biomedical image processing,and plasmonics.

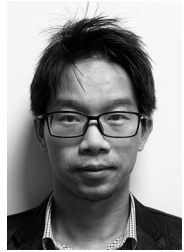

**Quan Wang** received the M.S. degree in optics from Xi'an Technological University, China, in 2018. He was then appointed as an associate engineer at a hi-tech company in Singapore. He is currently working toward a Ph.D. degree in Biomedical Engineering at the University of Strathclyde. His current research interests include fluorescence lifetime imaging microscopy systems (FLIM), flow cytometry, diffuse correlation spectroscopy (DCS), and biomedical image processing.

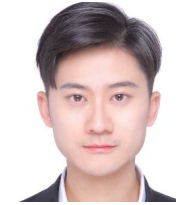

**Ziao Jiao** Ziao Jiao is currently working toward the PhD degree in biomedical engineering from the University of Strathclyde. His research interests include optical microscope and biomedical imaging. He is trying to develop different kinds of portable microscopes with artificial intelligence function for low-resource-limited areas and real-time disease detection.

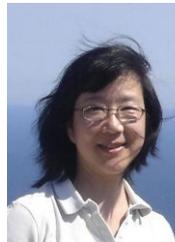

**Yu Chen** received the Ph.D. degree in nanoscale physics from Birmingham University, Birmingham, U.K., in 2000. She was then appointed to a lectureship in molecular nanometrology at the University of Strathclyde, Glasgow, U.K.,in 2007 funded by an EPSRC Science and Innovation Award. She was a senior lecturer from 2013 to 2017 and became a Reader in 2018.

Her main research activities lie in the creation and characterization of nanoscale structures for their unique physical and chemical properties utilizing optical and electron microscopy and spectroscopy. Her recent research focuses on fluorescent nanoclusters, surface plasmon enhanced effects, including two-photon luminescence, energy transfer and SERRS from noble metal nanoparticles, arrays, and porous media, with strong links to biomedical imaging and sensing, as well as nanoparticle-cell interaction and cytotoxicity. Other work includes 3D visualization of nanoparticles, stability of Au nanoparticles, and creating nanostructures using chemical lithography and direct writing.

She is a fellow of the Royal Microscopical Society, a member of the European Microscopy Society, the Institute of Physics, and the American Chemical Society. She serves as an expert panel member for the European Cooperation in Science and Technology, a referee for top international journals, an invited speaker and a committee member for a number of international conferences and workshops.

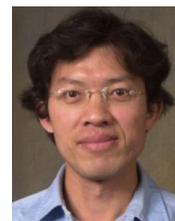

**David Day-Uei Li** (Member, IEEE) received my Ph.D. degree in electrical engineering from National Taiwan University, Taipei, Taiwan, in 2001. He then joined the Industrial Technology Research Institute in Taiwan, working on CMOS optical and wireless communication chipsets. From 2007 to 2011, he worked at the University of Edinburgh, Edinburgh, on two European projects focusing on CMOS single-photon avalanche diode sensors and systems. He then took the lectureship in biomedical engineering at the University of Sussex, Brighton, in mid-2011, and in 2014 he joined the University of Strathclyde, Glasgow, as a Senior Lecturer. He has authored more than 100 journal and conference papers and holds 12 patents. His research interests include CMOS sensors and systems, mixed signal circuits, embedded systems, optical communications, FLIM systems and analysis, and field-programmable gate array/graphics processing unit computing. His research exploits advanced sensor technologies to reveal low-light but fast biological phenomena.